\documentclass{WileyMSP-template}

\usepackage{array}
\usepackage{mdwmath}
\usepackage{mdwtab}
\usepackage{eqparbox}
\usepackage{adjustbox}
\usepackage{booktabs}
\usepackage{color}
\begin{document}

\pagestyle{fancy}

\title{In-situ dielectric Al$_2$O$_3$/$\beta$-Ga$_2$O$_3$ Interfaces Grown Using Metal-organic Chemical Vapor Deposition}

\maketitle

\author{Saurav Roy,*}
\author{Adrian E. Chmielewski,}
\author{Arkka Bhattacharyya,}
\author{Praneeth Ranga,}
\author{Rujun Sun,}
\author{Michael A. Scarpulla,}
\author{Nasim Alem, and}
\author{Sriram Krishnamoorthy*}



\begin{affiliations}
S. Roy, A. Bhattacharyya, P. Ranga, Dr. R. Sun, Prof. M. A. Scarpulla, Prof. S. Krishnamoorthy \\
Department of Electrical and Computer Engineering, University of Utah, SLC, UT 84112\\
Dr. A. E. Chmielewski, Prof. N. Alem\\
Department of Materials Science and Engineering, Pennsylvania State University, University Park, Pennsylvania 16802\\
Email Address: saurav.roy@utah.edu, sriram.krishnamoorthy@utah.edu
\end{affiliations}



\begin{abstract}
\justify
High quality dielectric-semiconductor interfaces are critical for reliable high performance transistors. We report the in-situ metalorganic chemical vapor deposition (MOCVD) of Al$_2$O$_3$ on $\beta$-Ga$_2$O$_3$ as a potentially better alternative to the most commonly used atomic layer deposition (ALD). The growth of Al$_2$O$_3$ is performed in the same reactor as Ga$_2$O$_3$ using trimethylaluminum and O$_2$ as precursors without breaking the vacuum at a growth temperature of 600 $^0$C. The fast and slow near interface traps at the Al$_2$O$_3$/ $\beta$-Ga$_2$O$_3$ interface are identified and quantified using stressed capacitance-voltage (CV) measurements on metal oxide semiconductor capacitor (MOSCAP) structures. The density of shallow and deep level initially filled traps (D$_{it}$) are measured using ultra-violet (UV) assisted CV technique. The average D$_{it}$ for the MOSCAP is determined to be 7.8 $\times$ 10$^{11}$  cm$^{-2}$eV$^{-1}$. The conduction band offset of the Al$_2$O$_3$/ Ga$_2$O$_3$ interface is also determined from CV measurements and found out to be 1.7 eV which is in close agreement with the existing literature reports of ALD Al$_2$O$_3$/ Ga$_2$O$_3$ interface. The current-voltage characteristics are also analyzed and the average breakdown field is extracted to be approximately 5.8 MV/cm. This in-situ Al$_2$O$_3$ dielectric on $\beta$-Ga$_2$O$_3$ with improved dielectric properties can enable  Ga$_2$O$_3$-based high performance devices.
\end{abstract}


\vspace{0.5cm}

\justify
The rise of $\beta$-Ga$_2$O$_3$ as a promising material for power electronic applications due to its extremely high breakdown field (8 MV/cm)\textsuperscript{\cite{pearton2018review, stepanov2016gallium}} and Baliga figure of merit (BFOM) has also provided impetus for the search of an appropriate insulator material as the gate dielectric for device applications. For transistor applications, a gate dielectric should exhibit low leakage currents, have low interface trap densities to achieve a controllable threshold voltage (V$_T$), and should also have a higher breakdown field compared to the underlying semiconductor. Many insulators such as SiO$_2$,\textsuperscript{\cite{zeng2016interface, zeng2017temperature, 8141864, 7904635}} Al$_2$O$_3$,\textsuperscript{\cite{hung2014energy, biswas2019enhanced}} SiN,\textsuperscript{\cite{carey2019comparison}} and HfO$_2$\textsuperscript{\cite{carey2019comparison, moser2017high}} etc. have been studied as a gate oxide material and passivation layers for Gallium Oxide devices. Extreme permittivity materials such as Barium Titanate (BaTiO$_3$) were also studied and used as dielectric material in $\beta$-Ga$_2$O$_3$ based transistors and heterojunction schottky barrier diodes.\textsuperscript{\cite{9274352, xia2019metal}} Complex oxides such as ternary rare earth alloys (Y$_{0.6}$Sc$_{0.4}$)$_2$O$_3$ were also investigated by Masten et. al. as gate dielectric for $\beta$-Ga$_2$O$_3$ based MOS structures.\textsuperscript{\cite{8700500}}

\vspace{0.5cm}

\par Among all the dielectric materials, Al$_2$O$_3$ is studied and used most extensively for $\beta$-Ga$_2$O$_3$ based devices due to its compatibility with $\beta$-Ga$_2$O$_3$. Chabak et. al. demonstrated various lateral MOSFETs with Al$_2$O$_3$ as gate dielectric showing excellent field strength.\textsuperscript{ \cite{8457153, chabak2016enhancement, 7470552}} Li et. al. also demonstrated $\beta$-Ga$_2$O$_3$ based vertical FinFET structures with outstanding Figure of Merit (FOM) using Al$_2$O$_3$ as gate dielectric.\textsuperscript{\cite{8993526, 8901439, 9046425}} Jian et. al. studied the effects of post deposition annealing on the interface trap density of the ALD deposited Al$_2$O$_3$ on (001) $\beta$-Ga$_2$O$_3$.\textsuperscript{\cite{jian2020deep}} In all these reported works the deposition of Al$_2$O$_3$ is primarily performed using ex-situ technique such as Atomic layer deposition (ALD), where after its growth, the gallium oxide substrate or epilayer is taken to a different reactor for dielectric deposition. During such a transfer step, it is hard to completely remove any interface contamination. Using in-situ deposition of Al$_2$O$_3$ in the same reactor as the underlying Ga$_2$O$_3$, the surface of the Ga$_2$O$_3$ can be kept pristine without exposure to the ambient. The quality of the dielectric could also improve due to the high temperature growth by metal-organic chemical vapor deposition (MOCVD) (500 to 1000 $^0$C), compared to ALD (100 to 300 $^0$C). MOCVD also enables high deposition rates, significantly higher than that of the ALD technique. MOCVD reactor can also facilitate high temperature in-situ annealing which can reduce the dielectric/semiconductor interface state density as well as the improve the bulk dielectric quality. The in-situ growth of Al$_2$O$_3$ on both Ga-face and N-face GaN has been studied extensively using MOCVD.\textsuperscript{\cite{liu2014metalorganic, liu2013fixed, liu2013situ, liu2014situ, 8051081}} In this paper, we demonstrate for the first time in-situ growth of Al$_2$O$_3$ using MOCVD and electrical characterization of metal/in-situ Al$_2$O$_3$/$\beta$-Ga$_2$O$_3$ vertical MOSCAPs.

\begin{figure}[t]
\centering
\includegraphics[width=8in,height=6cm, keepaspectratio]{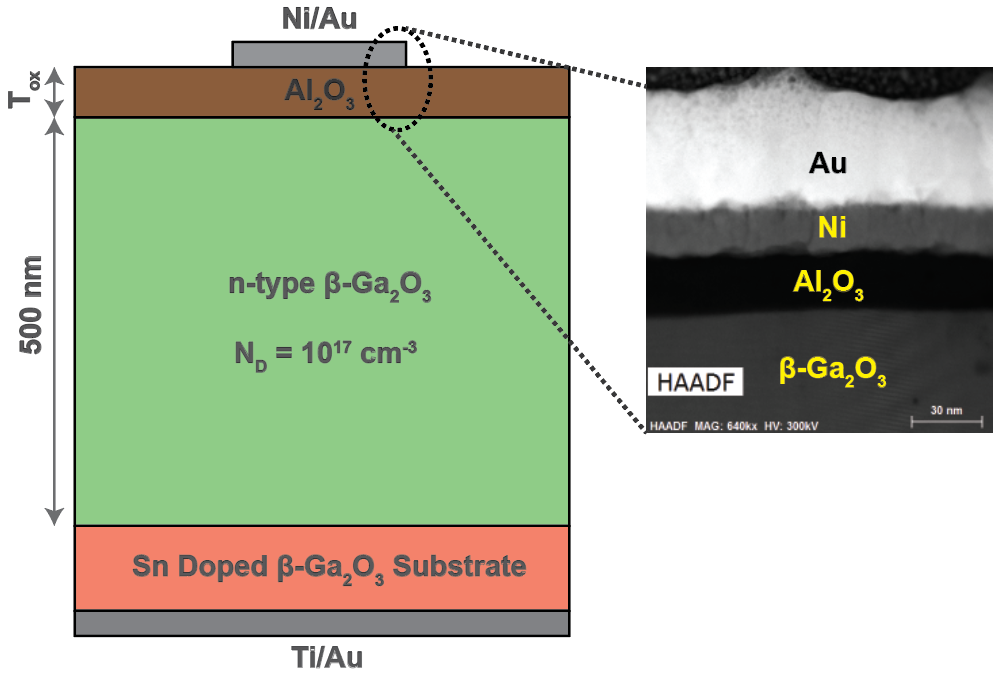}
\caption{Schematic diagram of the fabricated MOSCAP structures and the cross sectional TEM image of the expanded region showing different layers.}
\label{fig3}
\end{figure}

\begin{figure}[t]
\centering
\includegraphics[width=8in,height=6cm, keepaspectratio]{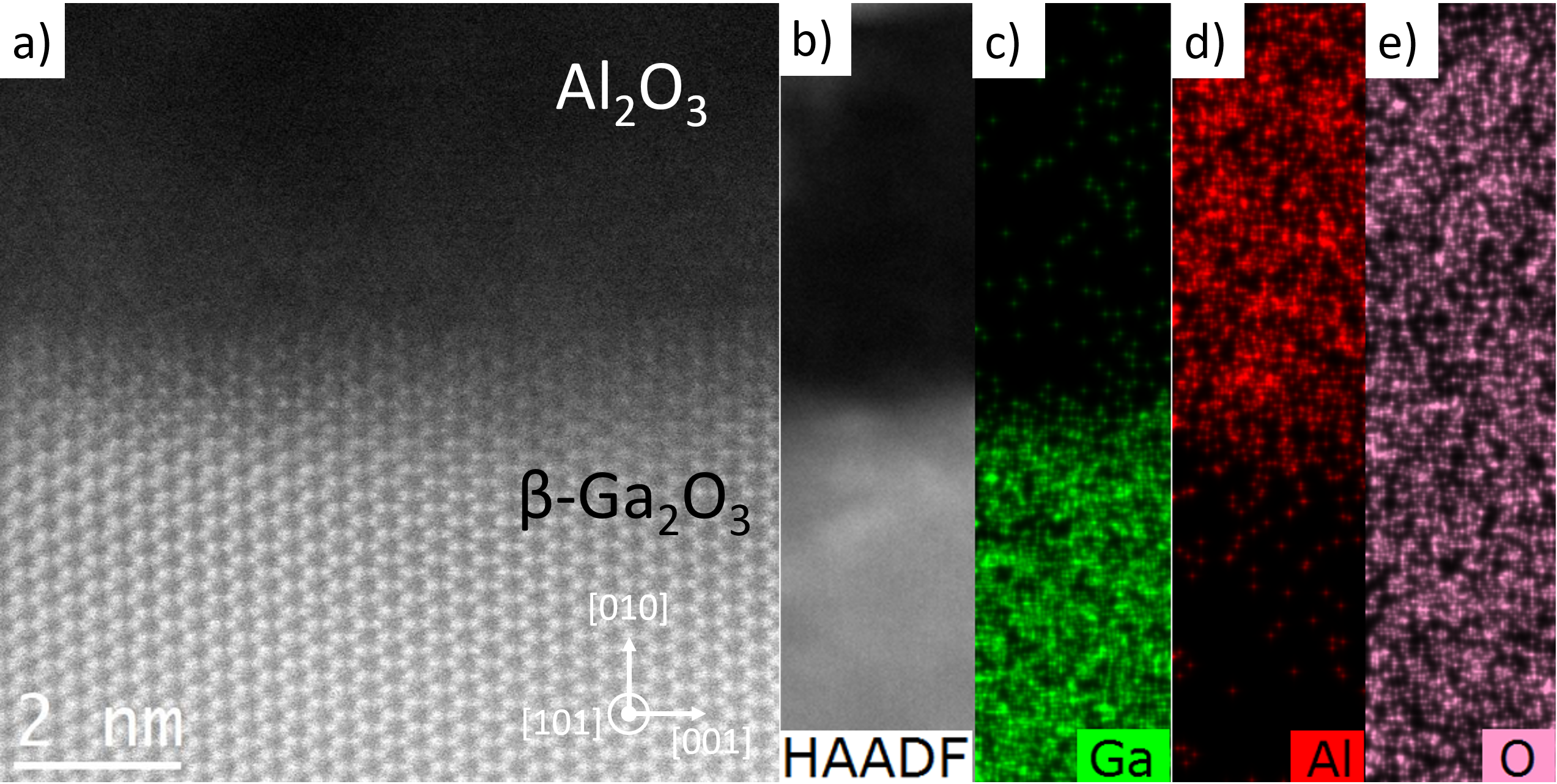}
\caption{(a) High resolution HAADF-STEM image of the Al$_2$O$_3$/$\beta$-Ga$_2$O$_3$ in the [101] direction. (b) low magnification HAADF-STEM image of the same region with its corresponding EDS maps of (c) Ga, (d) Al and (e) O.}
\label{fig1}
\end{figure}

\begin{figure}[t]
\centering
\includegraphics[width=8in,height=6cm, keepaspectratio]{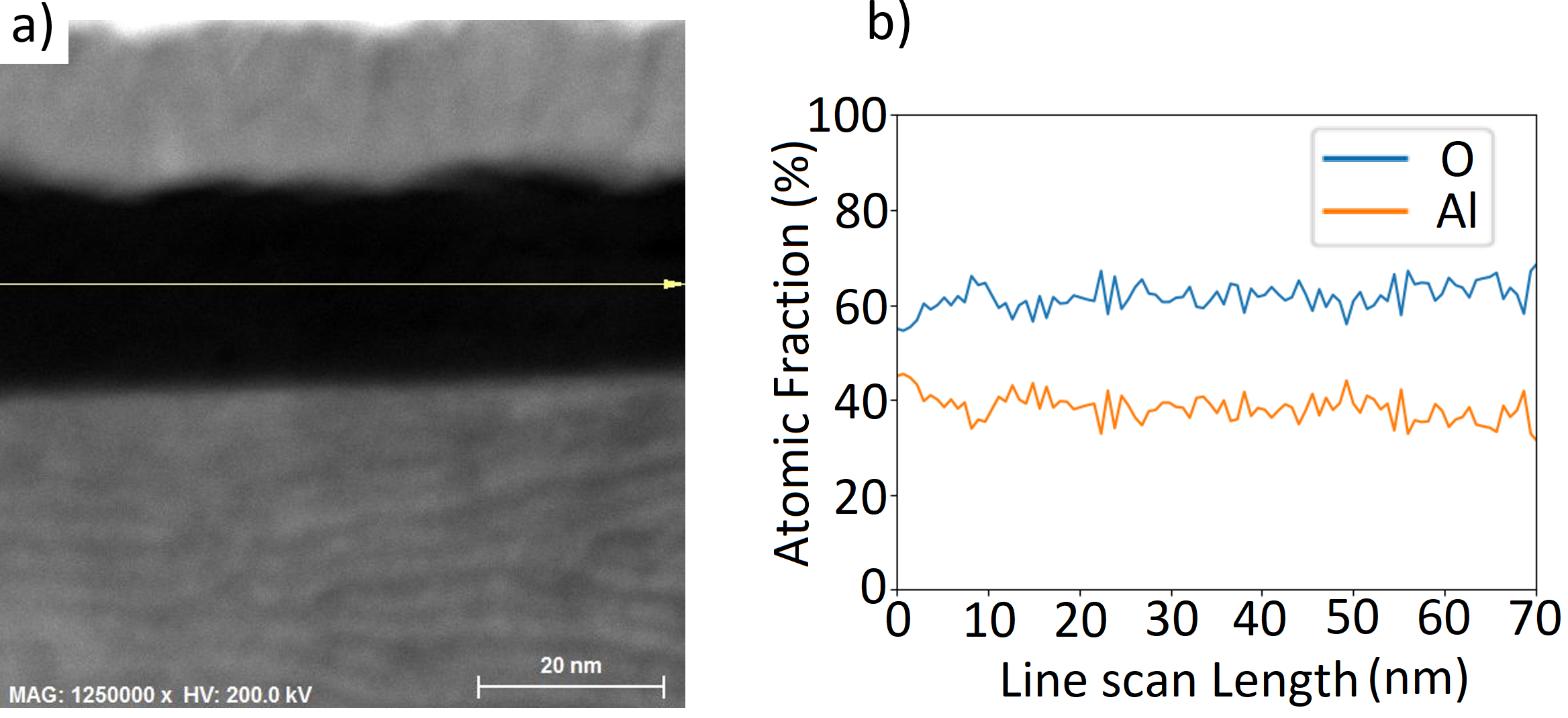}
\caption{(a) HAADF-STEM image of the Al$_2$O$_3$/$\beta$-Ga$_2$O$_3$ interface within the direction of the EDS line scan indicated by the yellow arrow. (b) Atomic fraction extracted form the EDS line scan.}
\label{fig2}
\vspace{-0.5cm}
\end{figure}

\vspace{0.5cm}

\vspace{0.5cm}

The growth of in-situ Al$_2$O$_3$ layer after the epitaxial growth of $\beta$-Ga$_2$O$_3$ on Sn doped (010) $\beta$-Ga$_2$O$_3$ and the fabrication of MOSCAPs as shown in Figure \ref{fig3} to analyze the in-situ Al$_2$O$_3$/$\beta$-Ga$_2$O$_3$ interface quality is discussed in the experimental section. To evaluate the abruptness of the Al$_2$O$_3$/$\beta$-Ga$_2$O$_3$ interface as well as the stoichiometry of the Al$_2$O$_3$, scanning transmission electron microscope imaging within a high angle annular dark field detector (HAADF-STEM) combined with energy dispersive x-ray spectroscopy (EDS) were used on a cross-section of the interface  prepared by focused ion beam (FEI Helios 660). HAADF-STEM images were performed using a FEI Titan G2 60-300 STEM with an acceleration voltage of 300kV, a condenser lens aperture of 70 µm and a probe current of about 100 pA. Figure \ref{fig1} (a) shows a HAADF-STEM image of the Al$_2$O$_3$/$\beta$-Ga$_2$O$_3$ interface in the [101] projection in which the $\beta$-Ga$_2$O$_3$  substrate is brighter then the Al$_2$O$_3$ film due to the heavier Ga atoms. The Al$_2$O$_3$/$\beta$-Ga$_2$O$_3$ interface is found to be abrupt with a transitional region of about 2 nm in which Ga atoms contrast vanishes and the high resolution is lost. The abruptness of the Al$_2$O$_3$/$\beta$-Ga$_2$O$_3$ interface  was also confirmed by EDS mapping (Bruker) performed in the same region. Figure \ref{fig1} (b) shows a HAADF-STEM image of the interface with its corresponding EDS maps of (c) Ga, (d) Al and (e) O. A smooth transition between the two materials is observed with a small diffusion of Al atoms at the top of the $\beta$-Ga$_2$O$_3$ epilayer. In addition, the composition of Al$_2$O$_3$ could be extracted from the EDS mapping analysis using the Cliff-Lorimer ratio technique detailed in the supporting information.\textsuperscript{\cite{cliff1975quantitative}} As a result, the film reveals an uniform composition with 38$\pm$2 \% of Al and 62$\pm$2 \% of O and stays relatively homogeneous along the Al$_2$O$_3$ film as shown by the EDS linescan of Figure \ref{fig2} (b).

\vspace{0.5cm}

To characterize the trap densities, the fabricated MOSCAPs with different thicknesses were subjected to a series of CV measurements using Keithley 4200, shown in Figure \ref{fig2}. The alternating current (AC) bias and frequency used were 30 mV and 1 MHz respectively and the MOSCAPs were swept at a rate of 200 mV/sec. The dielectric constant was extracted to be 8.2-8.5 from the accumulation capacitance on all three samples. The methodology used to extract the trap densities is same as the technique developed by Liu et. al. \textsuperscript{\cite{liu2013fixed}} and presented briefly here. The MOSCAPs were first taken from depletion (D) to accumulation (A) and kept at accumulation for 10 minutes to fill all the empty electron traps. After 10 minutes, they were again swept back to depletion (1st sweep). Another CV curve is measured by sweeping the MOSCAP from depletion to accumulation and immediately back from accumulation to depletion (2nd sweep). The total charge (Q$_I$) within the MOSCAPs is calculated from the equation (\ref{eq1}),

\begin{figure}[t]
\centering
\includegraphics[width=8in,height=12cm, keepaspectratio]{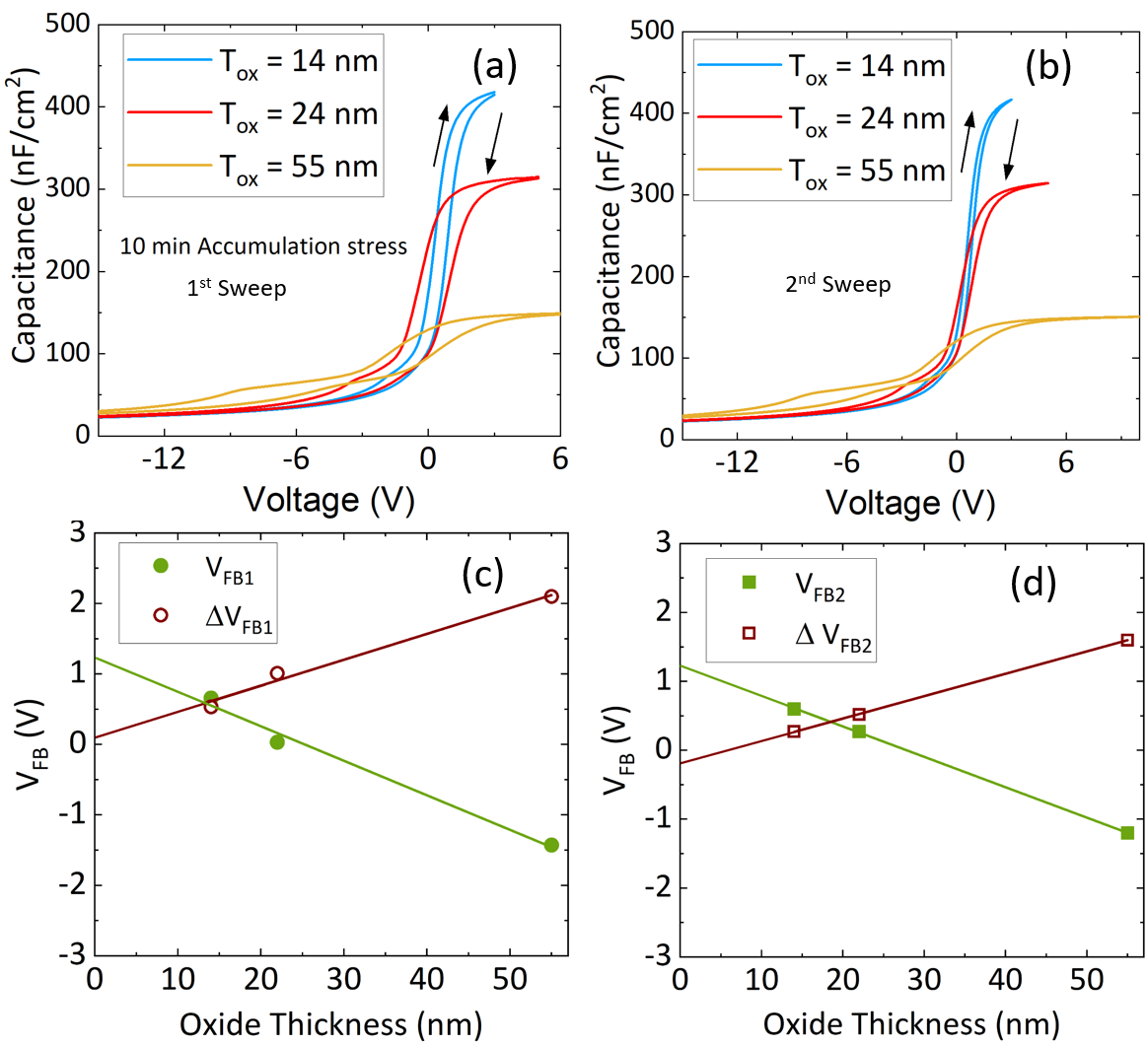}
\caption{(a) First $\&$ (b) second sweep of the CV hysteresis plots for MOSCAPs with three different dielectric thicknesses. The corresponding Flat band voltage vs oxide thickness for (c) first $\&$ (d) second sweeps. }
\label{fig4}
\end{figure}

\begin{equation}
\label{eq1}
V_{FB} = -\frac{qQ_IT_{ox}}{\epsilon \epsilon_0} + V_{FB}^{Ideal}
\end{equation}

where, V$_{FB}$ is the flat band voltage, T$_{ox}$ is the oxide thickness, V$_{FB}^{Ideal}$ is the ideal flat band voltage, $q$ is the electron charge, $\epsilon$ is the relative permittivity of the Al$_2$O$_3$ layer, and $\epsilon_0$ is the absolute permittivity. Assuming all the charges are located at or near the oxide/semiconductor interface, Q$_I$ can be determined from the linear fit of the V$_{FB}$ vs oxide thickness plot. The change in V$_{FB}$ for D to A and A to D sweeps of the hysteresis plot can be attributed to the total amount of initially empty trapped charge within the MOSCAPs as described by the equation (\ref{eq2}),

\begin{equation}
\label{eq2}
\Delta V_{FB} = -\frac{q\Delta Q_TT_{ox}}{\epsilon \epsilon_0}
\end{equation}

Where, $\Delta$V$_{FB}$ is the flat band voltage difference between the D to A and A to D sweeps and $\Delta$Q$_T$ is the trapped electron density. Figure \ref{fig4} (a) and (b) shows the 1st and 2nd sweeps of the CV hysteresis plots respectively for the MOSCAPs with three different Al$_2$O$_3$ thicknesses. The CV measurements were repeatable after the 2nd sweep, overlapping with the 2nd sweep measurements. Different forward bias voltages were applied for MOSCAPs with different Al$_2$O$_3$ thicknesses to maintain approximately same electric field through the oxide ($\sim$ 2 MV/cm). Figure \ref{fig4} (c) and (d) shows the corresponding V$_{FB}$ and $\Delta$V$_{FB}$ for the 1st and second sweeps. Flat band voltages were extracted from the deflection point of the capacitance voltage plot.\textsuperscript{\cite{winter2013new}} V$_{FB1}$ and V$_{FB2}$ are the flat band voltages calculated from the D to A sweeps of the 1st and 2nd sweep, respectively. Fixed charges (Q$_{I1}$ $\&$ Q$_{I2}$) are determined from the slope of the flat band voltage vs oxide thickness plot. Q$_{I1}$ is the fixed charge of the MOSCAP without any bias history. So, Q$_{I2}$ is the total fixed near interface charge in the same MOSCAP, but also contains trapped electrons due to the measurement sweeps and 10 minutes of accumulation stress. Now, $\Delta$V$_{FB1}$ and $\Delta$V$_{FB2}$ are the flat band voltage differences, between each pair of D to A and A to D sweeps of the 1st and 2nd sweep respectively. During the 10 minutes of accumulation stress of the first hysteresis measurement electrons fill the empty slow near interface states and due to the high emission time of the slow traps these electrons remain trapped afterwards. However, electron will fill the fast near interface states also during the 10 min accumulation time. So, $\Delta$V$_{FB1}$ can be attributed to the initially empty fast and slow near interface states ($\Delta$Q$_{T1}$). Only the fast traps will respond during the second hysteresis and subsequent measurements since the slow traps were filled during the first sweep. So $\Delta$V$_{FB2}$ can be attributed to only the fast near interface states($\Delta$Q$_{T2}$). Density of the initially empty slow near interface traps can be evaluated from the difference between $\Delta$Q$_{T1}$ and $\Delta$Q$_{T2}$ ($\left| \Delta Q_{T1} -\Delta Q_{T2} \right|$). However, during the first D to A and A to D sweeps, there could be electron emission from some of the slow near interface states due to the applied reverse bias. So, the difference in flat band voltage for the D to A sweeps of 1st and 2nd sweep can be attributed to the more stably filled slow near interface states ($\Delta$Q$_{TS}$). All the fixed and trapped charges described here are summarized in Table 1. The positive fixed charge observed in this work is similar to the previous report of ALD Al$_2$O$_3$/ (-201) Ga$_2$O$_3$ interface.\textsuperscript{\cite{hung2014energy}}
\vspace{-0.1 cm}
\begin{center}
    
\begin{table}
 \caption{Fixed charge and trap densities at the Al$_2$O$_3$/$\beta$-Ga$_2$O$_3$ interface}
  \begin{tabular}[htbp]{@{}llllll@{}}
    \hline
    Q$_{I1}$ & Q$_{I2}$ & $\Delta$Q$_{T1}$  & $\Delta$Q$_{T2}$ & $\left| \Delta Q_{T1} -\Delta Q_{T2} \right|$ & $\Delta$Q$_{TS}$ \\
    \hline
    2.3 $\times$ 10$^{12}$ cm$^{-2}$  & 2 $\times$ 10$^{12}$ cm$^{-2}$  & -1.7 $\times$ 10$^{12}$ cm$^{-2}$ & -1.2 $\times$ 10$^{12}$ cm$^{-2}$ & 5 $\times$ 10$^{11}$ cm$^{-2}$ & 3 $\times$ 10$^{11}$ cm$^{-2}$  \\
    \hline
  \end{tabular}
\end{table}

\end{center}

The conduction band offset at the Al$_2$O$_3$/$\beta$-Ga$_2$O$_3$ heterostructure can be evaluated using equation (\ref{eq3})  as V$_{FB}$ can be written as,\textsuperscript{\cite{hung2014energy}}

\begin{equation}
\label{eq3}
q V_{FB} = -qE_{ox}T_{ox} + (\phi_b - \Delta E_C - \phi_S)
\end{equation}

where, E$_{ox}$ is the electric field in the oxide at flat band condition due to the fixed charge at the, $\phi_b$ is Ni/Al$_2$O$_3$ conduction band barrier height, $\phi_S$ the difference between the conduction band and fermi level in $\beta$-Ga$_2$O$_3$, and $\Delta$E$_C$ is the conduction band offset between $\beta$-Ga$_2$O$_3$ and Al$_2$O$_3$. $\phi_b$ was considered to be 3 eV as evaluated by Zhang et. al. using internal photoemission spectroscopy.\textsuperscript{\cite{zhang2014direct}} From Figure \ref{fig4} (c) and (d), ($\phi_b - \Delta E_C - \phi_S$) is extracted to be 1.23 eV.  Using equation \ref{eq3}, the value of $\Delta$E$_C$ is calculated to be 1.7 eV which matches well with the previous reports on the ALD Al$_2$O$_3$/$\beta$-Ga$_2$O$_3$ heterostructure.

 \begin{figure}[t]
\centering
\includegraphics[width=8in,height=8cm, keepaspectratio]{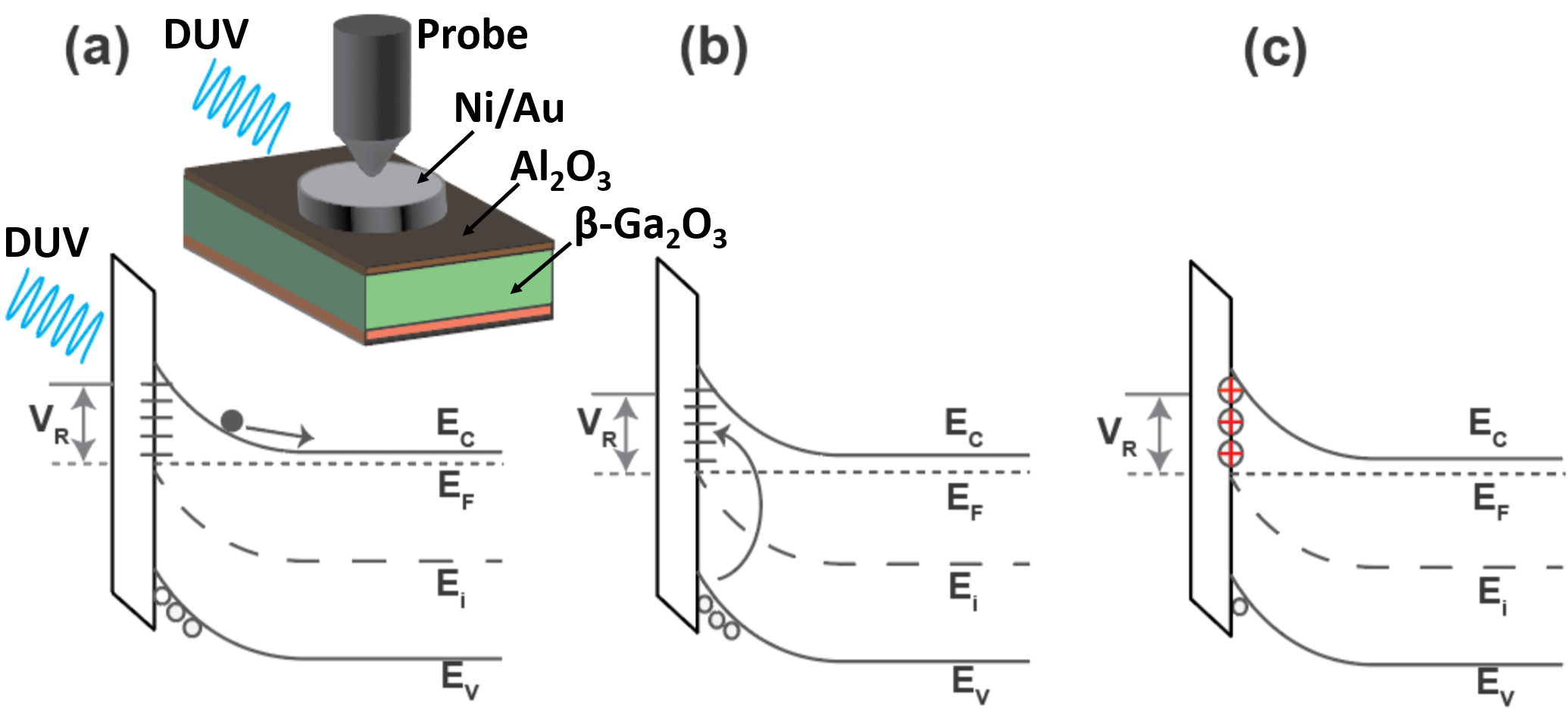}
\caption{Representative energy band diagram for the UV assisted CV measurements showing (a) generation of electron-hole pairs when excited by UV rays, (b) generated holes recombining with the trapped electrons at the interface, and (c) trap states became positively charged after the electron hole recombination process.}
\label{fig5}
\vspace{-0.5cm}
\end{figure}

\vspace{0.5cm}
 Now, to find the density of all the initially filled slow near interface states and the trap distribution as a function of energy dependence near the $\beta$-Ga$_2$O$_3$ band edge, we use UV assisted CV technique developed by Swenson et. al.\textsuperscript{\cite{swenson2009photoassisted}} and later modified by Liu et. al.\textsuperscript{\cite{liu2020improved}} This technique was also used by Jian et. al.\textsuperscript{\cite{jian2020deep}} for ALD Al$_2$O$_3$/(001) $\beta$-Ga$_2$O$_3$ MOSCAPs and briefly summarized here. The MOSCAPs were first taken from D to A and kept at accumulation for 10 minutes to fill all the interface states. After 10 minutes time the MOSCAPs were again swept back to depletion without any optical excitation. This dark curve is considered as the ideal CV curve, assuming all the states are filled with electrons. Next the MOSCAPs were held at depletion and were exposed to UV light using a DUV LED with 254 nm wavelength for 5 minutes. After 5 minutes the UV light is turned off and the MOSCAPs were held at depletion for 15 minutes followed by another sweep from D to A. The dark curve is then shifted horizontally to match the post DUV CV curve and obtained as the reference CV curve. The total density of the all the interface states is calculated from the difference between the post DUV and the reference CV curve ($\Delta$V) and their energy dependence is calculated from the amount of band bending (dielectric/semiconductor surface potential) that occurs during the bias sweep.
 
  \begin{figure}[t]
\centering
\includegraphics[width=8in,height=10cm, keepaspectratio]{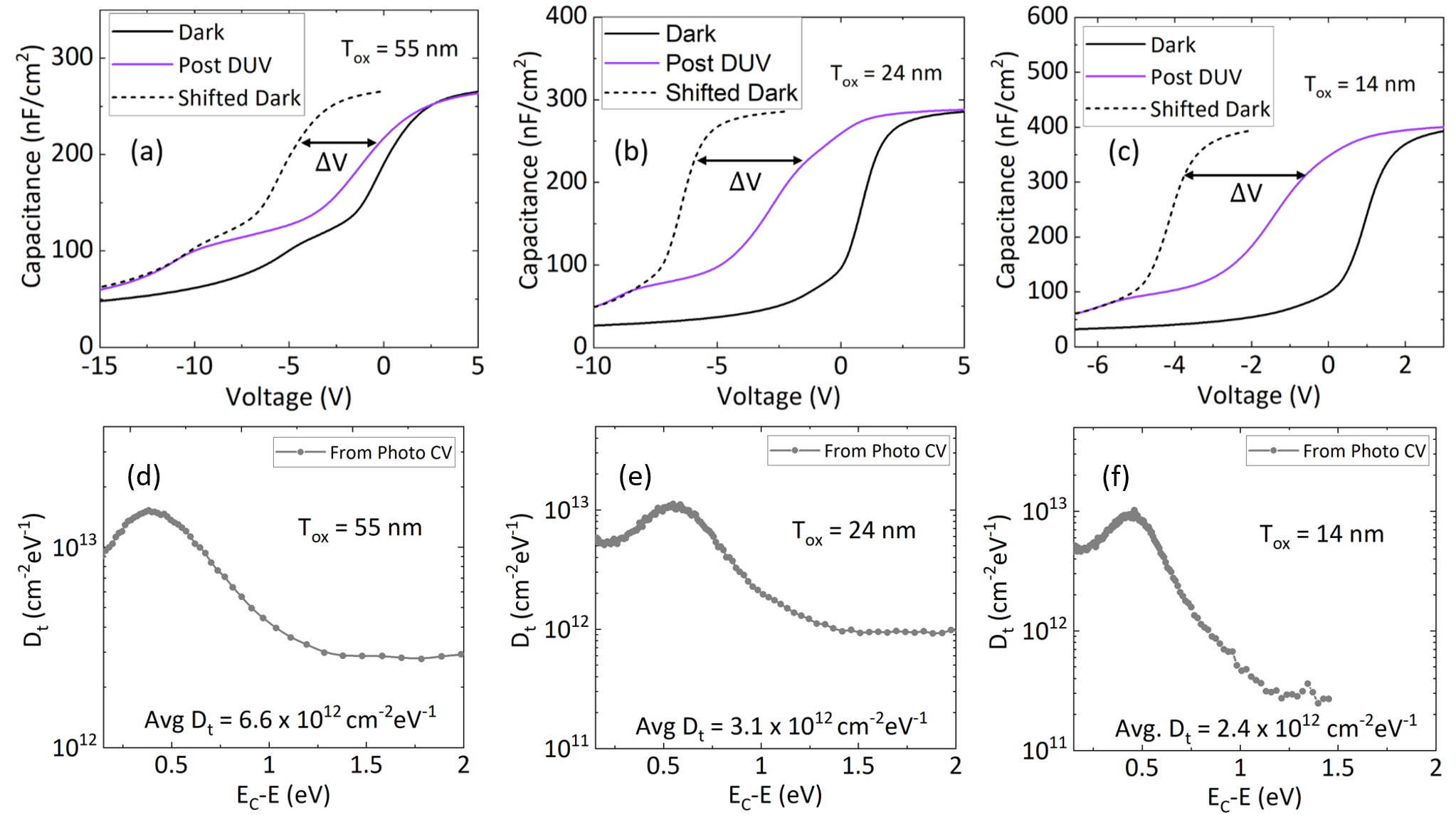}
\caption{UV assisted CV characteristics and trap density vs energy distribution plots for MOSCAPs with (a) $\&$ (d) 55 nm, (b) $\&$ (e) 24 nm, (c) $\&$ (f) 14 nm Al$_2$O$_3$ thickness. }
\label{fig6}
\vspace{-0.5cm}
\end{figure}
 
 \vspace{0.5cm}
 
Figure \ref{fig5} (a) to (c) shows energy band diagrams of the MOSCAPs under applied bias and UV illumination. During the UV exposure, electron hole pairs are generated and the photo-generated holes either recombine with the electrons trapped in the interface or gets self-trapped, which results in the ledge observed in the post DUV curve. Due to the recombination of photo generated holes with the trapped electrons at the Al$_2$O$_3$/$\beta$-Ga$_2$O$_3$ interface, the trapped states become positively charged, which changes the width of the depletion region. Some of the extra holes also get trapped in the dielectric bulk hole traps inside the Al$_2$O$_3$ layer. Figure \ref{fig6} (a) to (c) shows CV plots for the MOSCAPs with three different oxide thicknesses. The black curve (Dark) is the A to D sweep after the 10 minutes of accumulation stress. The purple curve is the D to A sweep after DUV exposure and the dashed black curve is the reference curve obtained after shifting the dark curve to match the post DUV plot. $\Delta$V in Figure \ref{fig6} (a) to (c) includes the effects of both interface traps at the Al$_2$O$_3$/$\beta$-Ga$_2$O$_3$ interface and bulk hole traps inside the Al$_2$O$_3$ layer. Density of all the initially filled interface and bulk dielectric traps is calculated using the equation (\ref{eq4}),

\begin{equation}
\label{eq4}
D_t = \frac{C_{ox}}{2A}\frac{d\Delta V}{d\psi_S}
\end{equation}

where, $\psi_S$ is the surface potential at the Al$_2$O$_3$/$\beta$-Ga$_2$O$_3$ interface, given by $\psi_S$ = $\frac{q\epsilon \epsilon_0N_DA^2}{2C_{dep}^2}$. N$_D$ is the doping density, $A$ is the area of the contacts, C$_{dep}$ is the depletion capacitance and C$_{ox}$ is the oxide capacitance.

\vspace{0.5cm}
Figure \ref{fig6} (d) to (f) shows the D$_t$ vs the trap energy distribution for the MOSCAPs with three different oxide thicknesses. Due to the existence of a positive valence band offset, the extra holes (which did not recombine with the interface traps) get accumulated at the Al$_2$O$_3$/$\beta$-Ga$_2$O$_3$ interface (some holes also get self-trapped). The rise in the D$_t$ near the conduction band edge can be attributed to the accumulated holes, either at the Al$_2$O$_3$/$\beta$-Ga$_2$O$_3$ interface or possibly also due to the self-trapped holes in the depletion region.\textsuperscript{\cite{yeluri2013capacitance}} The average D$_t$ for each of the MOSCAPs are obtained after eliminating the contribution from holes by neglecting the rise in D$_t$. \par

 \vspace{0.5cm}

We can also see from Figure \ref{fig6} (d) to (f) that the average D$_t$ reduces as the thickness of the Al$_2$O$_3$ layer decreases. There would not be any dielectric thickness dependence if all the traps were located at the Al$_2$O$_3$/$\beta$-Ga$_2$O$_3$ interface. Instead, due to the presence of bulk hole traps inside the Al$_2$O$_3$ layer, some of the holes get captured inside the Al$_2$O$_3$ layer. So, the density of the interface states (D$_{it}$) without the contribution from the bulk hole traps in the Al$_2$O$_3$ is calculated from the equation (\ref{eq5}).\textsuperscript{\cite{liu2020improved}}

\begin{equation}
\label{eq5}
D_t = D_{it} + T_{ox}\frac{dn_{ht}}{2d\psi_S}
\end{equation}

\begin{figure}[t]
\centering
\includegraphics[width=8in,height=6cm, keepaspectratio]{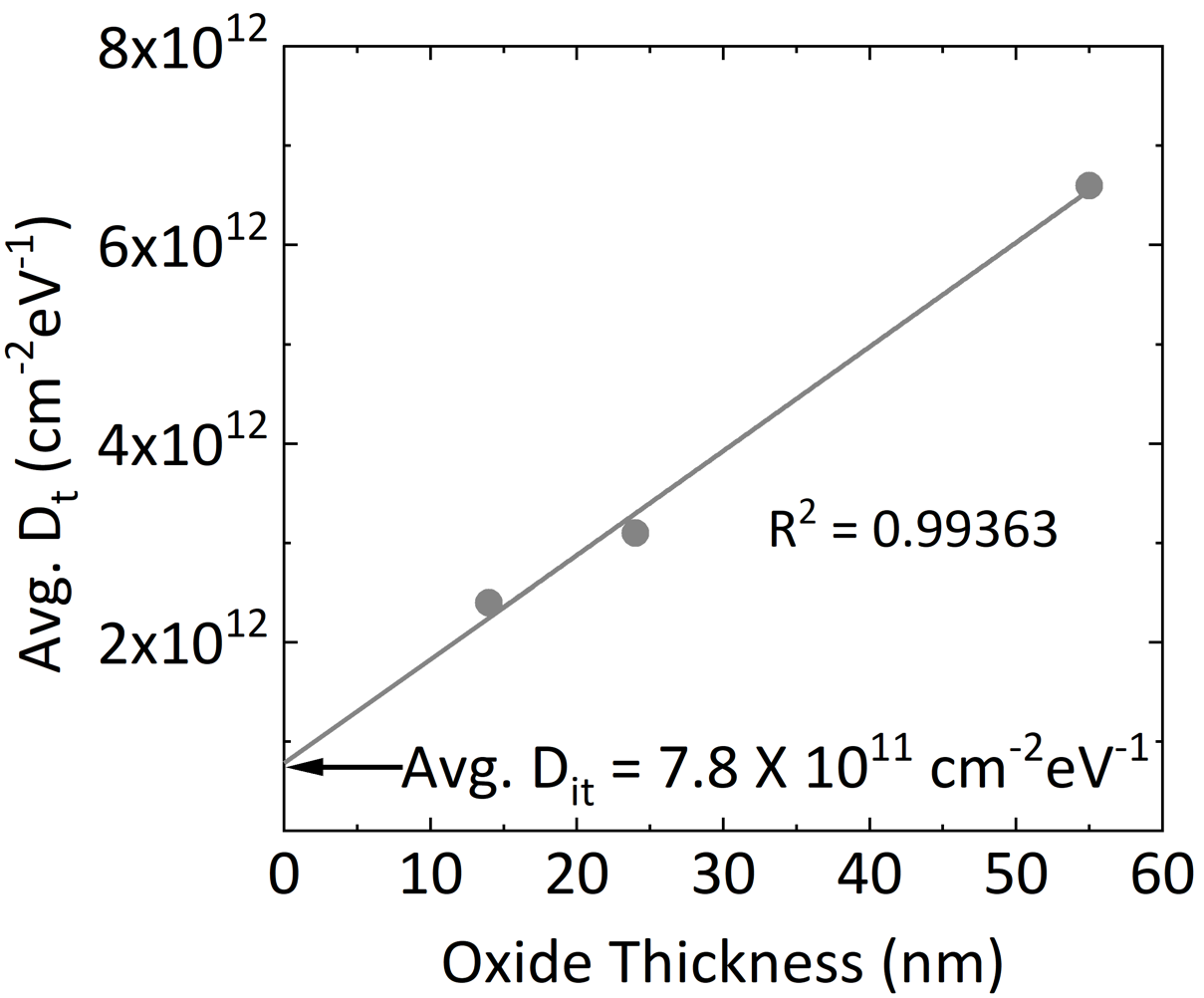}
\caption{Average trap density vs oxide thickness plot and the corresponding linear fit.}
\label{fig7}
\end{figure}

Where, n$_{ht}$ is the amount of bulk hole traps inside the Al$_2$O$_3$. So, average D$_{it}$ is obtained to be 7.8 $\times$ 10$^{11}$  cm$^{-2}$ eV$^{-1}$ by extrapolating the linear fit of average D$_t$ vs oxide thickness for T$_{ox}$ = 0 as shown in Figure \ref{fig7}. The amount of bulk hole traps (n$_{ht}$) is estimated from the  equation \ref{eq5} to be 2.1 $\times$ 10$^{17}$  cm$^{-3}$ eV$^{-1}$. The fact that we see a linear relationship of D$_t$ vs oxide thickness clearly indicates that not all the holes are self-trapped in the $\beta$-Ga$_2$O$_3$ and in-fact can move and get inside the Al$_2$O$_3$ layer and get trapped.  The mobility of holes could be significant enough to cause bulk hole trapping in the dielectric as also extracted by Akyol \textsuperscript{\cite{akyol2020simulation}} for $\beta$-Ga$_2$O$_3$ schottky barrier diode based photodetector. The value of D$_{it}$ obtained for the in-situ Al$_2$O$_3$ MOSCAPs are significantly lower than the previous reports of ALD Al$_2$O$_3$/ (001) $\beta$-Ga$_2$O$_3$ interface and may have resulted from the O vacancies and Al dangling bonds at the interface. 

\vspace{0.5cm}

\begin{figure}[t]
\centering
\includegraphics[width=8in,height=5.2cm, keepaspectratio]{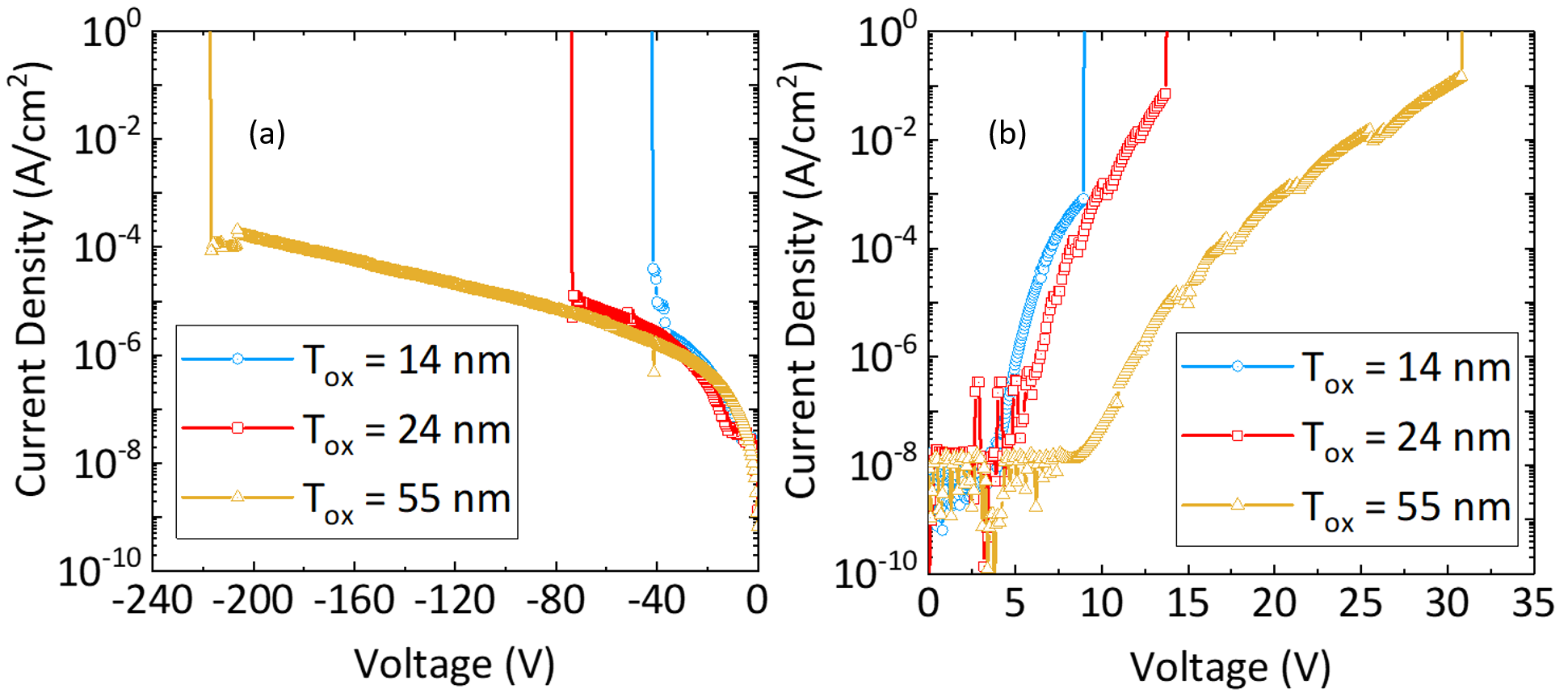}
\caption{(a) Reverse and (b) forward current voltage characteristics for MOSCAPs with three different dielectric thicknesses.}
\label{fig8}
\vspace{-0.5cm}
\end{figure}

\begin{figure}[t]
\centering
\includegraphics[width=8in,height=5.5cm, keepaspectratio]{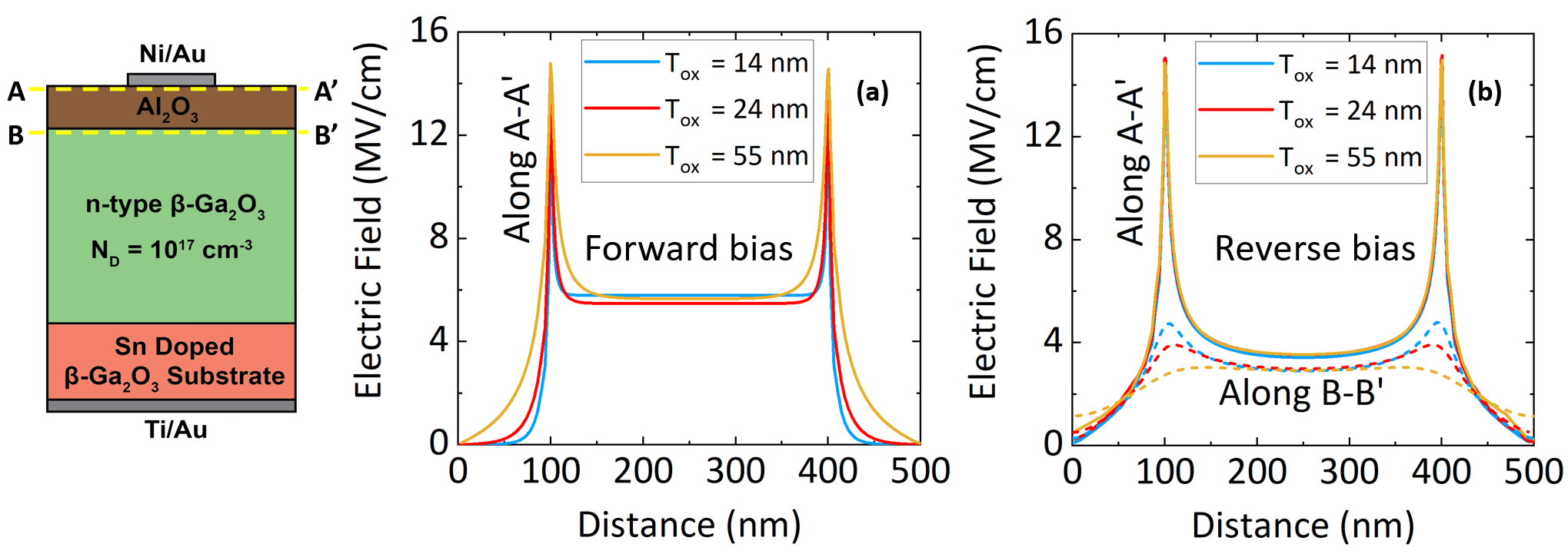}
\caption{Simulated electric field profiles for the MOSCAPs with three different Al$_2$O$_3$ thicknesses at (a) forward breakdown volatge along A-A'(9, 14, and 32 V for T$_{ox}$ = 14, 24, and 55 nm respectively) and (b) reverse breakdown volatge along A-A' (solid line) and B-B' (dashed line) (-40, -80, and -200 V for T$_{ox}$ = 14, 24, and 55 nm respectively). Electric field is plotted only at the Al$_2$O$_3$ region at forward bias because entire bias voltage is dropped across the oxide layer during accumulation}
\label{fig9}
\vspace{-0.5cm}
\end{figure}

\vspace{0.5cm}
To determine leakage currents and breakdown voltages of the MOSCAPs, reverse and forward current voltage (IV) characteristics is measured and shown in Figure \ref{fig8} (a) and (b) respectively. The measurement is performed at a sweep rate of 200 mV/sec. Breakdown voltages with positive bias on the gate metal (accumulation) for MOSCAPs with 14, 24 and 55 nm Al$_2$O$_3$ thicknesses are 9, 14, and 32 V, respectively. The breakdown field for all the MOSCAPs were determined using Sentaurus device simulator\textsuperscript{\cite{sentaurus}} and the field profile is shown in Figure \ref{fig9} (a). For the accumulation case with positive potential on the gate, the entire voltage is dropped across the Al$_2$O$_3$ layer and so the breakdown can be attributed to the breakdown of the dielectric. The average breakdown field at the center of the gate metal region and peak field at the edges of the metal for the Al$_2$O$_3$ dielectric are estimated to be $\sim$ 5.8 MV/cm and 15 MV/cm respectively from electric field simulations as shown in Figure \ref{fig9} (a). Electric fields were shown only in the Al$_2$O$_3$ region during forward bias as the entire bias voltage is dropped across the oxide layer. The barrier to tunneling for the MOSCAPs were also estimated using trap assisted tunneling model and extracted to be 1.1 eV and presented in the supporting information. The reverse breakdown voltage from the IV characteristics in Figure \ref{fig8} (a) for MOSCAPs with 14, 24 and 55 nm Al$_2$O$_3$ thicknesses are approximately 40, 80, and 200 V, respectively. From the electric field distribution in Figure \ref{fig9} (b), a peak field of $\sim$ 15 MV/cm is observed at the metal edges but the center of the gate metal, the electric field is less than the observed field in case of forward bias ($\sim$ 3.5 MV/cm as compared to 5.8 MV/cm). This is due to the fact that, at reverse bias, the total applied voltage is dropped across both Al$_2$O$_3$ and Ga$_2$O$_3$ region and a significant amount of electric field is present in the Ga$_2$O$_3$ ($\sim$ 2.6 MV/cm) as also shown by the dashed curves in Figure \ref{fig9} (b). At reverse bias, all the three MOSCAPs (T$_{ox}$ = 14, 24 and 55 nm) have a peak electric field of around $\sim$ 15 MV/cm, indicating that the breakdown is probably happening at the metal edges. These high edge electric fields can be reduced by creating a field plate structure using the same in-situ Al$_2$O$_3$ material showing the versatility of this technique.


 \vspace{0.5cm}
The in-situ Al$_2$O$_3$ grown using MOCVD, in this work, is found to have reduced fixed charges and interface traps (D$_{it}$) compared to the more commonly used ALD technique. A lower fixed charge concentration of +2 × 10$^{12}$  cm$^{-2}$ is found when compared to ALD Al$_2$O$_3$ on (-201) $\beta$-Ga$_2$O$_3$ (+3.6 $\times$ 10$^{12}$  cm$^{-2}$) reported by Hung et. al.\textsuperscript{\cite{hung2014energy}} A lower interface trap density (D$_{it}$) of 7.8 $\times$ 10$^{11}$  cm$^{-2}$eV$^{-1}$ is also found in comparison to ALD Al$_2$O$_3$/ (001) $\beta$-Ga$_2$O$_3$ interface reported by Jian et. al. (1.34 $\times$ 10$^{12}$  cm$^{-2}$eV$^{-1}$).\textsuperscript{\cite{jian2020deep}} It is to be noted that no high temperature post deposition annealing (PDA) or post metallization annealing (PMA) is performed on these MOSCAPs, which is expected to further improve the interface quality.\textsuperscript{\cite{chan2018first}} Further experiments to understand the effect of growth parameters such as growth temperature, chamber pressure and precursor ratios and also higher temperature PDA will be necessary to improve the dielectric constant, fixed charge, interface traps and critical breakdown fields of the Al$_2$O$_3$ layers in these fully MOCVD-grown Al$_2$O$_3$/$\beta$-Ga$_2$O$_3$ MOSCAPs. Also, detailed work is necessary to investigate the origin of the interface traps which may have resulted from native defects and impurities at the interface and can be performed using structural and spectroscopic techniques.




\vspace{0.5cm}

In conclusion, we report in-situ Al$_2$O$_3$ growth using MOCVD as a potentially better alternative to ALD and the interface quality is investigated using capacitance voltage characterization. The presence of near interface fixed charge in these MOSCAPs were identified and quantified using the linear relationship between flat band voltage and oxide thicknesses and a stable positive fixed charge of +2 $\times$ 10$^{12}$  cm$^{-2}$ is calculated. The sheet density of trap states with fast and slow emission times were also calculated from linear relationship of the CV hysteresis versus Al$_2$O$_3$ thickness plots. The density of all the near interface traps and their energy dependence were calculated using UV-assisted CV technique as low as +7.8 $\times$ 10$^{11}$  cm$^{-2}$. Conduction band offset using the linear relationship between flat band voltage and oxide thicknesses is calculated to be 1.7 eV. Furthermore, the average breakdown fields for the MOCVD grown Al$_2$O$_3$ material is found to be approximately 5.8 MV/cm. This approach of in-situ dielectric deposition on $\beta$-Ga$_2$O$_3$ can pave the way as gate dielectrics for future $\beta$-Ga$_2$O$_3$ based high performance MOSFETs due to its promise of improved interface and bulk quality compared to other conventional dielectric deposition techniques.
\vspace{-0.5cm}
\section*{Experimental Section}
The MOSCAP fabrication started with the growth of a 500 nm $\beta$-Ga$_2$O$_3$ epilayer on Sn-doped bulk (010) $\beta$-Ga$_2$O$_3$ substrates (NCT Japan) using Agnitron MOVPE reactor with far injection showerhead design. Prior to loading the sample in the growth chamber, the substrates were cleaned using acetone, methanol and DI water and followed by a 15 min dip in HF solution. Triethylgallium (TEGa) and O$_2$ were used as the precursors, Ar as the carrier gas and diluted silane as the dopant. The growth process of $\beta$-Ga$_2$O$_3$ was performed at a low temperature of 600 $^0$C and at a chamber pressure of 60 Torr.\textsuperscript{\cite{bhattacharyya2020low,ranga2020delta}} The gas flow values for TEGa and O$_2$ were set to 65 and 800 sccm and the Ar flow rate was set to 1100 sccm. A silane flow rate of 0.06 nmol/min was used to achieve a silicon doping concentration of 10$^{17}$  cm$^{-3}$. After the Ga$_2$O$_3$ growth, a 30 second purge step was used to purge all the unreacted precursors from the reactor. During this purge step, the reactor pressure was ramped down from 60 to 15 Torr whereas the temperature is kept same at 600 $^0$C. The O$_2$ flow rate was also reduced from 800 to 500 sccm during this 30 second purge step. Following this, growth time of the Al$_2$O$_3$ layer was varied to achieve dielectric thicknesses of 14, 24 and 55 nm, on three different substrates. In-situ Al$_2$O$_3$ was grown at a growth rate of 1.2 nm/min using the Trimethylaluminum (TMAl) precursor with a flow rate of 5.33 $\mu$mol/min and O$_2$ with a flow rate of 500 sccm. To find the thickness of the dielectric Al$_2$O$_3$ layer, the samples were patterned using photolithography and the Al$_2$O$_3$ dielectric layer was etched away using buffered oxide etch (BOE) with photoresist as the mask. The etch depth is measured using atomic force microscopy and the details are provided in the supporting information. After the growth of the Al$_2$O$_3$ layer, Ti (50 nm)/Au (100nm) ohmic contacts were sputter deposited on the back side of the MOSCAPs. Finally, circular Ni (10 nm)/Au (50 nm) gate contacts were patterned using standard photolithography and deposited on the Al$_2$O$_3$ surface using E-beam evaporation. 


\medskip
\textbf{Supporting Information} \par 
Supporting Information is available from the author.

\medskip
\textbf{Acknowledgements} \par 
This material is based upon work supported by the Air Force Office of Scientific Research under Award Number FA9550-18-1-0507 (Program Manager: Dr. Ali Sayir). Any opinions, findings, and conclusions or recommendations expressed in this material are those of the author(s) and do not necessarily reflect the views of the United States Air Force. The authors also acknowledge funding from the National Science Foundation (NSF) through Grant DMR-1931652. The work at Penn State University was supported by the Air Force Office of Scientific Research (AFOSR) program FA9550-18-1-0277 and GAME MURI, 10059059-PENN.

\medskip

%

\bibliographystyle{MSP}
\bibliography{Ref}

\begin{thebibliography}{10}
\providecommand{\url}[1]{\texttt{#1}}
\providecommand{\urlprefix}{URL }

\bibitem{pearton2018review}
S.~Pearton, J.~Yang, P.~H. Cary~IV, F.~Ren, J.~Kim, M.~J. Tadjer, M.~A. Mastro,
\newblock \emph{Applied Physics Reviews} \textbf{2018}, \emph{5}, 1 011301.

\bibitem{stepanov2016gallium}
S.~Stepanov, V.~Nikolaev, V.~Bougrov, A.~Romanov,
\newblock \emph{Rev. Adv. Mater. Sci} \textbf{2016}, \emph{44} 63.

\bibitem{zeng2016interface}
K.~Zeng, Y.~Jia, U.~Singisetti,
\newblock \emph{IEEE Electron Device Letters} \textbf{2016}, \emph{37}, 7 906.

\bibitem{zeng2017temperature}
K.~Zeng, U.~Singisetti,
\newblock \emph{Applied Physics Letters} \textbf{2017}, \emph{111}, 12 122108.

\bibitem{8141864}
K.~D. {Chabak}, J.~P. {McCandless}, N.~A. {Moser}, A.~J. {Green},
  K.~{Mahalingam}, A.~{Crespo}, N.~{Hendricks}, B.~M. {Howe}, S.~E. {Tetlak},
  K.~{Leedy}, R.~C. {Fitch}, D.~{Wakimoto}, K.~{Sasaki}, A.~{Kuramata}, G.~H.
  {Jessen},
\newblock \emph{IEEE Electron Device Letters} \textbf{2018}, \emph{39}, 1 67.

\bibitem{7904635}
A.~J. {Green}, K.~D. {Chabak}, M.~{Baldini}, N.~{Moser}, R.~{Gilbert}, R.~C.
  {Fitch}, G.~{Wagner}, Z.~{Galazka}, J.~{Mccandless}, A.~{Crespo}, K.~{Leedy},
  G.~H. {Jessen},
\newblock \emph{IEEE Electron Device Letters} \textbf{2017}, \emph{38}, 6 790.

\bibitem{hung2014energy}
T.-H. Hung, K.~Sasaki, A.~Kuramata, D.~N. Nath, P.~Sung~Park, C.~Polchinski,
  S.~Rajan,
\newblock \emph{Applied Physics Letters} \textbf{2014}, \emph{104}, 16 162106.

\bibitem{biswas2019enhanced}
D.~Biswas, C.~Joishi, J.~Biswas, K.~Thakar, S.~Rajan, S.~Lodha,
\newblock \emph{Applied Physics Letters} \textbf{2019}, \emph{114}, 21 212106.

\bibitem{carey2019comparison}
P.~H. Carey~IV, J.~Yang, F.~Ren, R.~Sharma, M.~Law, S.~J. Pearton,
\newblock \emph{ECS Journal of Solid State Science and Technology}
  \textbf{2019}, \emph{8}, 7 Q3221.

\bibitem{moser2017high}
N.~A. Moser, J.~P. McCandless, A.~Crespo, K.~D. Leedy, A.~J. Green, E.~R.
  Heller, K.~D. Chabak, N.~Peixoto, G.~H. Jessen,
\newblock \emph{Applied Physics Letters} \textbf{2017}, \emph{110}, 14 143505.

\bibitem{9274352}
N.~K. {Kalarickal}, Z.~{Feng}, A.~F.~M. {Anhar Uddin Bhuiyan}, Z.~{Xia},
  W.~{Moore}, J.~F. {McGlone}, A.~R. {Arehart}, S.~A. {Ringel}, H.~{Zhao},
  S.~{Rajan},
\newblock \emph{IEEE Transactions on Electron Devices} \textbf{2021},
  \emph{68}, 1 29.

\bibitem{xia2019metal}
Z.~Xia, H.~Chandrasekar, W.~Moore, C.~Wang, A.~J. Lee, J.~McGlone, N.~K.
  Kalarickal, A.~Arehart, S.~Ringel, F.~Yang, et~al.,
\newblock \emph{Applied Physics Letters} \textbf{2019}, \emph{115}, 25 252104.

\bibitem{8700500}
H.~N. {Masten}, J.~D. {Phillips}, R.~L. {Peterson},
\newblock \emph{IEEE Transactions on Electron Devices} \textbf{2019},
  \emph{66}, 6 2489.

\bibitem{8457153}
K.~D. {Chabak}, D.~E. {Walker}, A.~J. {Green}, A.~{Crespo}, M.~{Lindquist},
  K.~{Leedy}, S.~{Tetlak}, R.~{Gilbert}, N.~A. {Moser}, G.~{Jessen},
\newblock In \emph{2018 IEEE MTT-S International Microwave Workshop Series on
  Advanced Materials and Processes for RF and THz Applications (IMWS-AMP)}.
  \textbf{2018} 1--3.

\bibitem{chabak2016enhancement}
K.~D. Chabak, N.~Moser, A.~J. Green, D.~E. Walker~Jr, S.~E. Tetlak, E.~Heller,
  A.~Crespo, R.~Fitch, J.~P. McCandless, K.~Leedy, et~al.,
\newblock \emph{Applied Physics Letters} \textbf{2016}, \emph{109}, 21 213501.

\bibitem{7470552}
A.~J. {Green}, K.~D. {Chabak}, E.~R. {Heller}, R.~C. {Fitch}, M.~{Baldini},
  A.~{Fiedler}, K.~{Irmscher}, G.~{Wagner}, Z.~{Galazka}, S.~E. {Tetlak},
  A.~{Crespo}, K.~{Leedy}, G.~H. {Jessen},
\newblock \emph{IEEE Electron Device Letters} \textbf{2016}, \emph{37}, 7 902.

\bibitem{8993526}
W.~{Li}, K.~{Nomoto}, Z.~{Hu}, T.~{Nakamura}, D.~{Jena}, H.~G. {Xing},
\newblock In \emph{2019 IEEE International Electron Devices Meeting (IEDM)}.
  \textbf{2019} 12.4.1--12.4.4.

\bibitem{8901439}
W.~{Li}, K.~{Nomoto}, Z.~{Hu}, D.~{Jena}, H.~G. {Xing},
\newblock \emph{IEEE Electron Device Letters} \textbf{2020}, \emph{41}, 1 107.

\bibitem{9046425}
W.~{Li}, K.~{Nomoto}, Z.~{Hu}, D.~{Jena}, H.~G. {Xing},
\newblock In \emph{2019 Device Research Conference (DRC)}. \textbf{2019}
  209--210.

\bibitem{jian2020deep}
Z.~Jian, S.~Mohanty, E.~Ahmadi,
\newblock \emph{Applied Physics Letters} \textbf{2020}, \emph{116}, 24 242105.

\bibitem{liu2014metalorganic}
X.~Liu, S.~Chan, F.~Wu, Y.~Li, S.~Keller, J.~Speck, U.~Mishra,
\newblock \emph{Journal of crystal growth} \textbf{2014}, \emph{408} 78.

\bibitem{liu2013fixed}
X.~Liu, J.~Kim, R.~Yeluri, S.~Lal, H.~Li, J.~Lu, S.~Keller, B.~Mazumder,
  J.~Speck, U.~Mishra,
\newblock \emph{Journal of Applied Physics} \textbf{2013}, \emph{114}, 16
  164507.

\bibitem{liu2013situ}
X.~Liu, R.~Yeluri, J.~Kim, S.~Lal, A.~Raman, C.~Lund, S.~Wienecke, J.~Lu,
  M.~Laurent, S.~Keller, et~al.,
\newblock \emph{Applied Physics Letters} \textbf{2013}, \emph{103}, 5 053509.

\bibitem{liu2014situ}
X.~Liu, J.~Kim, D.~Suntrup, S.~Wienecke, M.~Tahhan, R.~Yeluri, S.~Chan, J.~Lu,
  H.~Li, S.~Keller, et~al.,
\newblock \emph{Applied Physics Letters} \textbf{2014}, \emph{104}, 26 263511.

\bibitem{8051081}
C.~{Gupta}, S.~H. {Chan}, A.~{Agarwal}, N.~{Hatui}, S.~{Keller}, U.~K.
  {Mishra},
\newblock \emph{IEEE Electron Device Letters} \textbf{2017}, \emph{38}, 11
  1575.

\bibitem{cliff1975quantitative}
G.~Cliff, G.~Lorimer,
\newblock \emph{Journal of Microscopy} \textbf{1975}, \emph{103}, 2 203.

\bibitem{winter2013new}
R.~Winter, J.~Ahn, P.~C. McIntyre, M.~Eizenberg,
\newblock \emph{Journal of Vacuum Science \& Technology B, Nanotechnology and
  Microelectronics: Materials, Processing, Measurement, and Phenomena}
  \textbf{2013}, \emph{31}, 3 030604.

\bibitem{zhang2014direct}
Z.~Zhang, C.~M. Jackson, A.~R. Arehart, B.~McSkimming, J.~S. Speck, S.~A.
  Ringel,
\newblock \emph{Journal of electronic materials} \textbf{2014}, \emph{43}, 4
  828.

\bibitem{swenson2009photoassisted}
B.~Swenson, U.~Mishra,
\newblock \emph{Journal of Applied Physics} \textbf{2009}, \emph{106}, 6
  064902.

\bibitem{liu2020improved}
W.~Liu, I.~Sayed, C.~Gupta, H.~Li, S.~Keller, U.~Mishra,
\newblock \emph{Applied Physics Letters} \textbf{2020}, \emph{116}, 2 022104.

\bibitem{yeluri2013capacitance}
R.~Yeluri, X.~Liu, B.~L. Swenson, J.~Lu, S.~Keller, U.~K. Mishra,
\newblock \emph{Journal of Applied Physics} \textbf{2013}, \emph{114}, 8
  083718.

\bibitem{akyol2020simulation}
F.~Akyol,
\newblock \emph{Journal of Applied Physics} \textbf{2020}, \emph{127}, 7
  074501.

\bibitem{sentaurus}
\emph{{Sentaurus Device User Manual, (Version N-2017.09, March 2017)}}.

\bibitem{chan2018first}
S.~H. Chan,
\newblock Ph.D. thesis, UC Santa Barbara, \textbf{2018}.

\bibitem{bhattacharyya2020low}
A.~Bhattacharyya, P.~Ranga, S.~Roy, J.~Ogle, L.~Whittaker-Brooks,
  S.~Krishnamoorthy,
\newblock \emph{Applied Physics Letters} \textbf{2020}, \emph{117}, 14 142102.

\bibitem{ranga2020delta}
P.~Ranga, A.~Bhattacharyya, A.~Chmielewski, S.~Roy, N.~Alem, S.~Krishnamoorthy,
\newblock \emph{Applied Physics Letters} \textbf{2020}, \emph{117}, 17 172105.

\end{thebibliography}


\newpage


\end{document}